\documentclass[prc,aps]{revtex4}
\usepackage{epsfig,amsfonts,amsmath,amssymb,amscd}

\begin{document}

\title{Neutrino energy loss rates and positron capture
rates on $^{55}$Co for presupernova and supernova physics}

\author{Jameel-Un Nabi\footnote{Corresponding author e-mail: jnabi00@gmail.com},
and  Muhammad Sajjad}

\affiliation{Faculty of Engineering Sciences, GIK Institute of
Engineering Sciences and Technology, Topi 23640, Swabi, NWFP,
Pakistan}
\begin{abstract}
Proton-neutron quasi-particle random phase approximation (pn-QRPA)
theory has recently being used for calculation of stellar weak
interaction rates of $fp$-shell nuclide with success. Neutrino
losses from proto-neutron stars play a pivotal role to decide if
these stars would be crushed into black holes or explode as
supernovae. The product of abundance and positron capture rates on
$^{55}$Co is substantial and as such can play a role in fine
tuning of input parameters of simulation codes specially in the
presupernova evolution. Recently we introduced our calculation of
capture rates on $^{55}$Co, in a luxurious model space of $7 \hbar
\omega$, employing the pn-QRPA theory with a separable
interaction. Simulators, however, may require these rates on a
fine scale. Here we present for the first time an expanded
calculation of the neutrino energy loss rates and positron capture
rates on $^{55}$Co on an extensive temperature-density scale.
These type of scale is appropriate for interpolation purposes and
of greater utility for simulation codes. The pn-QRPA calculated
neutrino energy loss rates are enhanced roughly up to two orders
of magnitude compared with the large-scale shell model
calculations and favor a lower entropy for the core of massive
stars. \vskip 0.1in\noindent\textbf{PACS} numbers: 21.60.Jz,
23.40.Bw, 26.30.Jk, 26.50.+x\vskip 0.2in
\end{abstract}
 \maketitle
\section{Introduction}

 Supernovae are nature's grandest explosions.
They are also responsible for synthesizing most of the elements of
nature, including those that form our own planet, Earth. Many
exotic states of matter, including black holes and neutron stars,
also owe their existence to supernovae. Since 1934, when Baade and
Zwicky [1] suggested that supernovae are energized by the collapse
of an ordinary star to a neutron star, scientists started debating
about the physical mechanism responsible for these spectacular
explosions (where the luminosity of the star becomes comparable to
that of an entire galaxy containing around $10^{11}$ stars$!$).
Whereas gravity remains the undisputed source of energy, the
relative roles of other physical phenomena including, but not
limited to, roles of neutrino, continue to be argued.

The evolution of a massive star (of masses above $10M_{\odot}$)
comprises of several stages including hydrogen, helium, carbon,
neon, oxygen and silicon burning. The time scales, temperatures,
densities, fuels, luminosities and neutrino losses for these
stages can be found in Ref. [2]. During the late phases of
evolution of these massive stars an iron core develops (of mass
around  $1.5 M_{\odot}$). The binding energy per nucleon curve
prohibits any further production of energy by nuclear fusion, yet
the neutrino losses continue unabated, exceeding the Sun's
luminosity by a factor of about $10^{15}$. Capture rates and
photodisintegration processes contribute in the lowering of the
degeneracy pressure required to counter the enormous self-gravity
force of the star. Under such extreme thermodynamic conditions,
neutrinos are produced in abundance. Eventually the collapse of
the iron core begins. The mechanism of core-collapse supernovae is
strongly believed to depend upon the transfer of energy from the
inner core to the outer mantle of the iron core. Neutrinos seem to
be the mediators of this energy transfer. The shock wave, produced
as a result, stalls due to photodisintegration and neutrino energy
losses. Once again the part played by neutrinos in this scenario
is far from being completely understood. Whereas Haxton [3]
proposed the mechanism of "preheating" by neutrinos as a means of
assistance for shock revival, Bruenn and Haxton [4] later
discouraged the preheating mechanism. They worked on two different
models, simulating weak and strong shock cases, and found out that
in neither case is the energy transferred to the matter by
neutrino-nucleus absorption significant in terms of preheating the
infalling iron-like material. More recently, Langanke and
collaborators [5] had some observations on the models of Bruenn
and Haxton [4] and reported much larger preshock heating rates,
albeit acting for too short a time to lead to consequences for
shock propagation. According to authors in Ref. [5] the inelastic
neutrino-nucleus scattering modifies the radiated neutrino spectra
and strongly reduces the high-energy spectral tail of the electron
neutrino burst at shock breakout.

A few milliseconds after the bounce, the proto-neutron star
accretes mass at a few tenths of solar mass per second. This
accretion, if continued even for one second, can change the
ultimate fate of the collapsing core resulting into a black hole.
Neutrinos are the main characters in this play and radiate around
10$\%$ of the rest mass converting the star to a neutron star
(e.g. [2]). Despite the small neutrino-nucleus cross sections, the
neutrinos flux generated by the cooling of a neutron star can
produce a number of nuclear transmutations as it passes the
onion-like structured envelope surrounding the neutron star.
Within $\sim$ 0.1 s of the beginning of the collapse the
nonthermal neutrino emission is dominated by electron neutrinos
owing to the decay and capture of leptons by nuclei and free
protons. The mean individual neutrino energies are some $\sim$ 10
MeV and constitute roughly 10$\%$ of the total available energy of
around $(3 - 5)$x $10 ^{53}$ erg [6].

The neutrino energy loss rates are important input parameters in
multi-dimensional simulations of the contracting proto-neutron
star. The reenergizing by charged-current electron neutrino and
antineutrino absorption on the dissociation-liberated protons and
neutrons in the postshock flow remains integral to the supernova
paradigm and neutrino transport is arguably the single most
important component of any supernova model [7]. (For a review of
supernova neutrino microphysics see also Ref. [8].) Parameter-free
multi-dimensional models, with neutrino transport included
consistently throughout the entire mass, yield conflicting results
on the key issue of whether the star actually explodes. Reliable
and microscopic calculations of neutrino loss rates and capture
rates can contribute effectively in the final outcome of these
simulations on world's fastest supercomputers.

During core infall electron neutrinos are produced almost entirely
by electron captures on free protons and nuclei and at
sufficiently high temperatures antineutrinos are also produced as
a result of positron captures on neutrons. Electron capture on
protons and positron capture on neutrons also play a crucial role
in the evolution of star and supernova explosion. During the
collapse and accretion phases, they decrease the degenerate
pressure in the stellar core. The neutrinos produced in these
capture processes carry the energy away and result in the lowering
of the entropy of the core. Positron captures are of great
importance in high temperature and low density locations. In such
conditions, a rather high concentration of positrons can be
reached from $e^{-} +e^{+} \leftrightarrow \gamma +\gamma $
equilibrium favoring the $e^{-} e^{+} $ pairs. Positron capture on
elements lying at the bottom of the valley of nuclear
beta-stability (so-called \textit{s} elements) may capture
positrons and be transformed into a proton-rich isobars (so-called
\textit{p} elements). The electron capture on proton and the
positron capture on neutron are considered important ingredients
in the modelling of Type-II supernovae [9].

Fuller, Fowler, and Newman (FFN) [10] performed the first-ever
extensive calculation of stellar weak rates including the capture
rates, neutrino energy loss rates and decay rates for a wide
density and temperature domain. They made this detailed
calculations for 226 nuclei in the mass range $21 \leq A \leq 60$.
They also stressed the importance of the Gamow-Teller (GT) giant
resonance strength in the capture of the electron and estimated
the GT centroids using zeroth-order ($0\hbar\omega$ ) shell model.
Later, Aufderheide et al. [11] extended the FFN work for heavier
nuclei with A $>$ 60. They tabulated the 90 top electron capture
nuclei averaged throughout the stellar trajectory for $0.40 \leq
Ye \leq 0.5$ (see Table. 25 therein). Since then theoretical
efforts were concentrated on the microscopic calculations of
capture rates of iron-regime nuclide. Large-scale shell model
(e.g. [12]) and the proton-neutron quasiparticle random phase
approximation theory (pn-QRPA) (e.g. [13]) were used extensively
and with relative success for the microscopic calculation of
stellar capture rates and neutrino energy losses. Monte Carlo
shell-model is an alternative to the diagonalization method and
allows calculation of nuclear properties as thermal averages (e.g.
[14]). However it does not allow for detailed nuclear
spectroscopy.

Nabi and Klapdor [15] calculated weak interaction rates for 709
nuclei with A = 18 to 100 in stellar matter using the pn-QRPA
theory. These included capture rates, decay rates, neutrino energy
loss rates, probabilities of beta-delayed particle emissions and
energy rate of these particle emissions. Since then these
calculations were further refined with use of more efficient
algorithms, incorporation of latest data from mass compilations
and experimental values, and fine-tuning of model parameters [16,
17, 18, 19].

$^{55}$Co is abundant in the presupernova conditions and is
thought to contribute effectively in the dynamics of presupernova
evolution. Aufderheide and collaborators [11] placed $^{55}$Co
among the list of top ten  most important capture nuclei during
the presupernova evolution. Later Heger et al. [20] also
identified $^{55}$Co as the most important nuclide for capture
purposes for massive stars ($25M_{\odot}$). Realizing the
importance of $^{55}$Co in astrophysical environments, Nabi,
Rahman and Sajjad [21] reported the calculation of electron and
positron capture rates on $^{55}$Co using the pn-QRPA theory (see
also Ref. [22]). However there was a need to perform a fine
calculation of these important capture rates on a detailed
temperature-density grid suitable for collapse simulation codes
(see, for example, Ref. [18, 23]). Due to the extreme conditions
prevailing in these scenarios, interpolation of calculated rates
within large intervals of temperature-density points posed some
uncertainty in the values of capture rates for collapse
simulators. Further, as mentioned above, the neutrino energy loss
rates needed to be included in these expanded calculations on a
detailed stellar temperature-density grid to make them more useful
in simulation codes.

In this paper we present for the first time an expanded
calculation of (anti)neutrino energy loss rates and positron
capture rates on $^{55}$Co at fine intervals of
temperature-density intervals. Section II deals with the formalism
of our calculation. Due to existing physical situation and lack of
experimental data the uncertainties present in stellar rate
calculations are considerable. We discuss the uncertainties of the
pn-QRPA model in Section III. In Section IV we will be presenting
some of our results. Comparisons with earlier calculations are
also included in this section. We finally will be concluding in
Section V and at the end Table V presents our expanded calculation
of (anti)neutrino energy loss rates and positron capture rates on
$^{55}$Co.

\section{The quasi-particle random phase approximation with a separable interaction}
The QRPA theory is an efficient way to generate GT strength
distributions. These strength distributions constitute a primary
and non-trivial contribution to the calculation of positron
capture and neutrino energy loss rates. Kar et al. [24] pointed
out that the quasiparticle random phase approximation (QRPA)
method is quite successful in predicting the weak interaction
rates of ground states all over the periodic table and also
stressed the need to extend these methods to non-zero temperature
domains relevant to presupernova and supernova conditions. QRPA is
also the method of choice in dealing heavy nuclei [25]. The QRPA
theory considers the residual correlations among the nucleons via
one particle one hole (1p-1h) excitations in a large model space.
Nabi and Klapdor [15] extended the QRPA model to configurations
more complex than 1p-1h.

We used the pn-QRPA theory to calculate the GT strength functions
and the associated capture and neutrino energy loss rates for
$^{55}$Co. The reliability of the pn-QRPA calculations was
discussed in detail by Nabi and Klapdor [13]. There the authors
compared the measured data (half lives and B(GT) strength) of
thousands of nuclide with the pn-QRPA calculations and got fairly
good comparison. We incorporated experimental data wherever
available to further strengthen the reliability of our calculated
rates. The calculated excitation energies (along with their log ft
values) were replaced with the experimental ones when they were
within 0.5 MeV of each other. Missing measured states were
inserted and inverse and mirror transitions were also taken into
account. We did not replace the theoretical levels with the
experimental ones beyond the excitation energy for which
experimental compilations had no definite spin and/or parity
assignment (2.98 MeV in case of $^{55}$Co). The pn-QRPA theory was
used with a separable interaction which granted us the liberty of
performing the calculations in a much larger single-particle basis
than a general interaction. We performed the pn-QRPA calculations
using a model space of seven major harmonic oscillator shells ($7
\hbar \omega$). The Hamiltonian for our calculations was of the
form
\begin{equation}
H^{QRPA} =H^{sp} +V^{pair} +V_{GT}^{ph} +V_{GT}^{pp},
\end{equation}
here $H^{sp}$ is the single-particle Hamiltonian, $V^{pair}$  is
the pairing force, $V_{GT}^{ph}$ is the particle-hole (ph)
Gamow-Teller force, and $V_{GT}^{pp}$  is the particle-particle
(pp) Gamow-Teller force. Single particle energies and wave
functions were calculated in the Nilsson model, which takes into
account nuclear deformations. Pairing was treated in the BCS
approximation. The proton-neutron residual interactions occurred
as particle-hole and particle-particle interaction. The
interactions were given separable form and were characterized by
two interaction constants $\chi$  and $\kappa$, respectively. The
selections of these two constants were done in an optimal fashion.
For details of the fine tuning of the Gamow-Teller strength
parameters, we refer to Ref. [26, 27]. In this work, we took the
values of $\chi = 0.2 MeV$ and $\kappa = 0.07 MeV$. Other
parameters required for the calculation of weak rates are the
Nilsson potential parameters, the deformation, the pairing gaps,
and the Q-value of the reaction. Nilsson-potential parameters were
taken from Ref. [28] and the Nilsson oscillator constant was
chosen as $\hbar \omega=41A^{-1/3}(MeV)$ (the same for protons and
neutrons). The calculated half-lives depend only weakly on the
values of the pairing gaps [29]. Thus, the traditional choice of
$\Delta _{p} =\Delta _{n} =12/\sqrt{A} (MeV)$ was applied in the
present work. The deformation parameter for $^{55}$Co, $\delta$,
was taken to be 0.06, according to M\"{o}ller and Nix [30]. (See
also the discussion on choice of deformation parameter in Ref.
[19].) Q-values were taken from the recent mass compilation of
Audi et al. [31].

The positron capture rates of a transition from the $ith$ state of
the parent to the $jth$ state of the daughter nucleus is given by
\begin{equation}
\lambda ^{^{pc} } _{ij} =\left[\frac{\ln 2}{D}
\right]\left[f_{ij}^{pc} (T,\rho ,E_{f} )\right]\left[B(F)_{ij}
+\left({\raise0.7ex\hbox{$ g_{A}  $}\!\mathord{\left/ {\vphantom
{g_{A}  g_{V} }} \right.
\kern-\nulldelimiterspace}\!\lower0.7ex\hbox{$ g_{V}  $}}
\right)^{2} B(GT)_{ij} \right].
\end{equation}
We took the value of D=6295s [32] and the ratio of the axial
vector to the vector coupling constant as -1.254 [33]. $B_{ij}'$s
are the sum of reduced transition probabilities of the Fermi B(F)
and GT transitions B(GT). Details of these reduced transition
probabilities can be found in Ref. [13, 17]. The phase space
integral $f_{ij}$ is an integral over total energy and for
positron capture it is given by
\begin{equation}
f_{ij}^{pc} \, =\, \int _{w_{l} }^{\infty }w\sqrt{w^{2} -1} (w_{m}
\, +\, w) ^{2} F(-Z,w)G_{+} dw.
\end{equation}
In above equation $w$ is the total energy of the electron
including its rest mass, $w_{l}$ is the total capture threshold
energy (rest+kinetic) for positron capture. F(-Z,w) are the Fermi
functions and were calculated according to the procedure adopted
by Gove and Martin [34]. G$_{+}$ is the Fermi-Dirac distribution
function for positrons.
\begin{equation}
G_{+} =\left[\exp \left(\frac{E+2+E_{f}
}{kT}\right)+1\right]^{-1},
\end{equation}
here $E = (w-1)$ is the kinetic energy of the positrons, $E_{f}$
is the Fermi energy of the positrons, $T$ is the temperature, and
$k$ is the Boltzmann constant.

The number density of electrons associated with protons and nuclei
is $\rho Y_{e}N_{A}$ ($\rho$ is the baryon density, $Y_{e}$ is
lepton to baryon ratio, and $N_{A}$ is Avogadro's number)
\begin{equation}
\rho Y_{e}=\frac{1}{\pi ^{2} N_{A} }(\frac{m_{e}c}{\hbar})^{3}\int
_{0}^{\infty }(G_{-}  -G_{+} )p^{2} dp,
\end{equation}
here $p = (w^{2}-1)^{1/2}$ is the positron momentum and Eqt. (5)
has the units of $mol \hspace{0.1in} cm^{-3}$. G$_{-}$ is the
Fermi-Dirac distribution function for electrons.
\begin{equation}
 G_{-} =\left[\exp \left(\frac{E-E_{f} }{kT}
 \right)+1\right]^{-1}.
\end{equation}

Eqt.5 was used for an iterative calculation of Fermi energies for
selected values of $Y_{e}$ and T. There is a finite probability of
occupation of parent excited states in the stellar environment as
result of the high temperature prevailing in the interior of
massive stars. Weak interactions then also have a finite
contribution from these excited states. The total positron capture
rate per unit time per nucleus is given by
\begin{equation}
\lambda_{pc} =\sum _{ij}P_{i} \lambda _{ij}^{pc}.
\end{equation}
The summation over all initial and final states was carried out
until satisfactory convergence in the rate calculations was
achieved. Here $P_{i}$ is the probability of occupation of parent
excited states and follows the normal Boltzmann distribution.

The neutrino energy loss rates can occur through four different
weak-interaction mediated channels: electron and positron
emissions, and, continuum electron and positron captures. The
neutrino energy loss rates were calculated using the same
formalism described above except that the phase space integral was
replaced by
\begin{equation}
f_{ij}^{\nu} \, =\, \int _{l }^{w_{m}}w\sqrt{w^{2} -1} (w_{m} \,
 -\, w)^{3} F(\pm Z,w)(1- G_{\mp}) dw,
\end{equation}
and by
\begin{equation}
f_{ij}^{\nu} \, =\, \int _{w_{l} }^{\infty }w\sqrt{w^{2} -1}
(w_{m} \,
 +\, w)^{3} F(-Z,w)G_{+} dw.
\end{equation}
For the decay channel Eqt. 8 was used for the calculation of phase
space integrals. Upper signs were used for the case of electron
emissions and lower signs for the case of positron emissions.
Regarding the capture channels, Eqt. 9 was used for the
calculation of phase space integrals keeping upper signs for
continuum electron captures and lower signs for continuum positron
captures.

The total neutrino energy loss rate per unit time per nucleus is
given by
\begin{equation}
\lambda_{\nu} =\sum _{ij}P_{i} \lambda _{ij}^{\nu},
\end{equation}
where $\lambda_{ij}^{\nu}$ is the sum of the electron capture and
positron decay rates for the transition $i \rightarrow j$.

On the other hand the total antineutrino energy loss rate per unit
time per nucleus is given by
\begin{equation}
\lambda_{\bar{\nu}} =\sum _{ij}P_{i} \lambda _{ij}^{\bar{\nu}},
\end{equation}
where $\lambda_{ij}^{\bar{\nu}}$ is the sum of the positron
capture and electron decay rates for the transition $i \rightarrow
j$.

\section{Uncertainties of the pn-QRPA model}
The uncertainties involved in stellar rate calculations are
considerable. The prevailing extreme physical conditions and model
parameters invoke uncertainties in the calculation. Lack of
experimental data in this scenario deteriorates the situation. The
``electron capture direction'' can be explored experimentally by
(n,p) experiments whereas the ``beta minus decay direction'' can
be explored by the (p,n) reactions. There are a handful of other
experiments which have been used by many theorists to shape the
centroid and width of the GT strength. However these experimental
data are not enough to completely explore the domain of nuclei
which are interesting from astrophysical viewpoint. There is no
experimental data concerning GT strength distribution from parent
excited states. In the stellar environment, at high temperatures
and densities, there is a finite probability of occupation of
parent excited states and transitions from these excited states
are sometimes many orders of magnitude higher than transitions
from ground states [19]. The weak interaction rates are calculated
using
\begin{equation}
\lambda_{ij}=\frac{ln2}{D}f_{ij}B_{ij}.
\end{equation}
Here $i$ represents the parent excited states and $j$ the
daughter's. The first factor is a constant and the second are
phase space integrals which can be calculated relatively accurate.
It is the third factor (the reduced transition probabilities)
which contains interesting nuclear physics and incorporates
uncertainties in the model. Again the reduced transition
probabilities is a sum of Fermi and GT component (see Eqt. 2).
Whereas the calculation of Fermi transition is rather
straightforward it is precisely the calculation of the excited
states and reduced transition probabilities of the GT transitions
which is the main cause of uncertainty of the underlying model.
The pn-QRPA model constructs parent and daughter excited states
and also calculates GT strength distribution among these states in
a microscopic fashion. In other words the Brink's hypothesis is
not employed in this calculation which increases the reliability
of the pn-QRPA calculations (Brink's hypothesis states that GT
strength distribution on excited states is identical to that from
ground state, shifted only by the excitation energy of the state).
As mentioned above there are still uncertainties present due to
the parameters of the model. Roughly, the parameters of the
pn-QRPA model can be divided into two different groups: (i)
"internal" parameters of the model which are by some means
adjustable (the pairing gaps and the GT strength parameters) and
(ii) "external" parameters for which input from other sources like
mass formulae (or experimental data, if available) is necessary
(these include single particle energies and wavefunctions,
deformations, Q values and neutron/proton separation energies).
Whereas "internal" parameters are of minor importance the
uncertainty in the "external" parameters must be viewed as the
limiting factor for the calculation of weak interaction rates of
unstable nuclide. The values taken for these parameters and their
optimal selection procedure were highlighted in the previous
section. In order to further increase the reliability of the
calculated rates experimental data were incorporated into the
model as also discussed in the previous section.

We, however, do have a reasonable amount of experimental data on
measured half-lives and as such the pn-QRPA theory was tested to
check the accuracy of the model against the experimentally known
half-lives using the \textit{same set} of parameters. The check
was performed in both "beta minus decay" and "electron capture"
directions.

In Tables (I) and (II), $N$ denotes the number of experimentally
known half-lives shorter than the limit in the second column, $n$
is the number (and percentage) of isotopes reproduced under the
condition given in the first column, and $\bar{x}$ is the average
deviation defined by
\begin{equation}
\bar{x} = \frac{1}{n}\sum_{i=1}^{n}x_{i},
\end{equation}
where\\
\hspace*{3.5cm}$x_{i}=T^{cal}_{1/2}/T^{exp}_{1/2}$ \hspace*{0.2cm}
if \hspace*{0.2cm} $T^{cal}_{1/2} \geq T^{exp}_{1/2}$\\
\vspace*{0.1cm}\\
\hspace*{3.5cm}$x_{i}=T^{exp}_{1/2}/T^{cal}_{1/2}$ \hspace*{0.2cm}
if \hspace*{0.2cm} $T^{cal}_{1/2} < T^{exp}_{1/2}$.\\
\vspace*{0.2cm}\\
($ T^{cal}_{1/2}$ is the calculated half-life using the pn-QRPA
model and $ T^{exp}_{1/2}$ is the corresponding measured
half-life.) For example, the pn-QRPA reproduces 93$\%$ of all
experimentally known half-lives shorter than 1 minute for
$\beta^{+}$/EC within a factor of 10 with an average deviation of
$\bar{x}$ = 1.718 and 95$\%$ of all known $\beta^{-}$-decaying
nuclei with half-lives less than a minute are reproduced within a
factor of 5 with an average deviation of $\bar{x}$ = 1.56. It can
be seen from the tables that the model works better with
increasing neutron excess (corresponding to shorter half-lives),
that is, with increasing distance from stability. This is in
agreement with the expectation, since forbidden transitions is
neglected in the calculation. This is also a promising feature
with respect to the prediction of unknown half-lives (specially
for unstable isotopes), implying that the predictions are made on
the basis of a realistic physical model (see also Table~I and
Table~J of Ref. [13] for predictive power of pn-QRPA in "electron
capture" direction, and Table~K of Ref. [13] for the predictive
power of the model in the "beta minus decay" direction).

\section{Results and comparison}
At temperatures pertinent to supernova environment we have a
finite probability of occupation of excited states and a
microscopic calculation of rates from these excited states is
desirable. Earlier, Nabi and Sajjad [19] did point to the fact
that the Brink's hypothesis ( and back resonances for calculation
of beta decay rates) is not a good approximation to use in stellar
rate calculations. Brink's hypothesis states that GT strength
distribution on excited states is identical to that from ground
state, shifted only by the excitation energy of the state whereas
the GT back resonances are states reached by the strong GT
transitions in the electron capture process built on ground and
excited states. The luxury of having a huge model space at our
disposal allowed us to perform the calculation of electron capture
rates from 30 excited states of $^{55}$Co. Table III shows the
calculated excited states of $^{55}$Co using the pn-QRPA theory.
For each parent state we calculated the GT strength distribution
in a microscopic fashion to around 200 excited states in daughter.

The GT strength distribution ($GT_{\pm}$) from ground state and
first two excited states of $^{55}$Co was presented earlier [21].
We calculated the position of our GT$_{-}$ (in this direction a
neutron is changed into a proton) centroid around 9.1 MeV for the
ground state. For the first two excited states of $^{55}$Co ($@$
2.2 MeV and 2.6 MeV), the corresponding centroids were placed at
9.2 MeV and 9.6 MeV, respectively. The total GT strength for
positron capture from ground state of $^{55}$Co was calculated to
be around 17.9. Table IV shows the calculated B(GT$_{-}$) strength
values for the ground state of $^{55}$Co. The strengths are given
up to energy of 10 MeV in daughter nucleus, $^{55}$Ni. Calculated
B(GT$_{-}$) strength of magnitude less than $10^{-3}$ are not
included in this table.

Recently, we presented the extensive calculation of electron
capture rates on  $^{55}$Co on a fine temperature-density scale
where we also discussed our results in detail [23]. Here we would
like to present a similar calculation for the positron capture and
the associated (anti)neutrino energy loss rates due to
weak-interaction mediated reactions in the core of massive stars.

Fig.1 shows four panels depicting our calculated positron capture
rates at selected temperature and density domain. The upper left
panel shows the positron capture rates in low-density region
($\rho [gcm^{-3}] =10^{0.5}, 10^{1.5}$ and $10^{2.5}$), the upper
right in medium-low density region ($\rho [gcm^{-3}] =10^{3.5},
10^{4.5}$ and $10^{5.5}$), the lower left in medium-high density
region ($\rho [gcm^{-3}] =10^{6.5}, 10^{7.5}$ and $10^{8.5}$) and
finally the lower right panel depicts our calculated positron
capture rates in high density region ($\rho [gcm^{-3}] =10^{9.5},
10^{10.5}$ and $10^{11}$). The positron capture rates are given in
logarithmic scales in units of $s^{-1}$. T$_{9}$ gives the stellar
temperature in units of $10^{9}$K. One should note the order of
magnitude differences in positron capture rates as the stellar
temperature increases. It can be seen from this figure that in the
low density region the positron capture rates, as a function of
stellar temperatures, are more or less superimposed on one
another. This means that there is no appreciable change in the
rates when increasing the density by an order of magnitude. We
also observe that the positron capture rates are almost the same
for the densities in the range $(10-10^{6})g/cm^{3}$. However as
we go from the medium high density region to high density region
these rates start to 'peel off' from one another. Orders of
magnitude difference in rates are observed (as a function of
density) in high density regions. When the densities increase
beyond the above stated range a decline in the positron capture
rate starts. For a given density the rates increase monotonically
with increasing temperatures.

Fig.2 and Fig.3 depict our calculated neutrino and antineutrino
energy loss rates due to $^{55}$Co. It is pertinent to mention
again that the neutrino energy loss rates (depicted in Fig.2)
contain contributions due to electron capture \textit{and}
positron decay on $^{55}$Co whereas the antineutrino energy loss
rates (Fig.3) are calculated due to contributions from positron
capture \textit{and} electron decay on $^{55}$Co. The energy loss
rates are given in logarithmic scales (in units of $MeV.s^{-1}$).
The figures again consist of four panels depicting the low,
medium-low, medium-high and high density domains for the core of a
massive star. We note the similarity between the positron capture
rates (Fig.1) and the antineutrino energy loss rates (Fig.3). The
later are slightly enhanced at the corresponding temperature and
density. On the other hand the neutrino energy loss rates exhibit
an entirely different pattern. We note that the neutrino energy
loss rates are orders of magnitude greater than the corresponding
antineutrino energy loss rates. This is an expected result (the
Q-value of $^{55}$Co for electron capture/positron decay is 3.452
MeV whereas the Q-value of $^{55}$Co in the other direction is
-8.692 MeV [31]).

The comparison of electron and positron capture rates on $^{55}$Co
with earlier calculations were presented in Ref.~[21]. Here we
present a comparison of the energy loss rates with the earlier
calculations. Fig.4 presents a comparison of our calculated
neutrino energy loss rates compared with large-scale shell model
[12] and FFN [10] calculations. The comparison is presented at
densities $(10^{3}, 10^{7}, 10^{11}) g.cm^{-3}$. Compared to
large-scale shell model results, our calculations lead to a larger
energy being carried away by the neutrinos and hence favor cooler
cores. We note that our corresponding numbers are roughly as big
as two orders of magnitude (at presupernova temperature and
density region). In high temperature and density regions our
calculated neutrino energy loss rates are in good comparison with
those of large-scale shell model. As far as comparison with the
pioneering work of FFN is concerned, we note that again our rates
are enhanced at presupernova temperature-density domain. However
at large stellar temperatures and densities, FFN neutrino energy
loss rates surpass our rates. There are two main reasons for this
enhancement of FFN rates. Firstly, FFN placed the centroid of the
GT strength at too low excitation energies in their compilation of
weak rates for odd-A and odd-odd nuclei [35]. Secondly, FFN
threshold parent excitation energies were not constrained and
extended well beyond the particle decay channel. At high
temperatures contributions from these high excitation energies
begin to show their cumulative effect. Simulators should take note
of our enhanced neutrino energy loss rates at the lower
temperatures and densities characteristic of the hydrostatic
phases of stellar evolution which may affect the temperature and
the corresponding lepton-to baryon ratio which becomes very
important going into stellar collapse.

The story is different for the comparison of antineutrino energy
loss rates (Fig.5). This time we note that the large-scale shell
model rates and FFN rates are much more enhanced compared to
pn-QRPA rate calculations. Nevertheless these are very small
numbers and can change by orders of magnitude by a mere change of
0.5 MeV, or less, in parent or daughter excitation energies and
are more indicative of the uncertainties present in the
calculation of the excitation energies.

Fig.6 shows our summed GT$_{-}$ strength as a function of excited
states in the daughter $^{55}$Ni. We note that almost all the
strength cumulates up to an energy around 12 MeV in $^{55}$Ni. No
appreciable strength is seen in $^{55}$Ni at higher excitation
energies.

We finally present our calculated positron capture rates, neutrino
and antineutrino energy loss rates on a detailed
temperature-density grid in Table V. Here Column 1 shows the
density in logarithmic scales (in units of $gcm^{-3}$), Column 2,
the stellar temperature in units of $10^{9}K$, Column 3 gives the
calculated positron capture rates in logarithmic scales (in units
of $sec^{-1}$) at the corresponding temperature and density
whereas Column 4 and Column 5 display the corresponding neutrino
and antineutrino energy loss rates again in logarithmic scales (in
units of $MeV.sec^{-1}$). All logarithms are taken to base 10.
Tables of rate calculations presented in earlier compilations
(e.g. Ref. [10, 12, 13, 36, 37]) were not presented on a detail
temperature-density grid and at times could lead to erroneous
results when interpolated. We hope that this table will prove more
useful for core-collapse simulators.

\section{Conclusions}
$^{55}$Co is advocated to play a key role amongst the iron-regime
nuclide controlling the dynamics of presupernova evolution of
massive stars. The capture rates and (anti)neutrino energy loss
rates on $^{55}$Co are used as nuclear physics input parameter for
multi-dimensional simulations. Reliable and detailed calculations
of these weak-interaction mediated rates are desirable for these
codes. These parameters may fine tune the final outcome of the
neutrino transport included multi-dimensional models.

Here we present, for the first time, an extensive calculation of
stellar positron capture rates and the (anti)neutrino loss rates
for $^{55}$Co on a fine temperature-density scale suitable for
simulation codes. According to authors in Ref. [11] and Ref. [20],
$^{55}$Co is a very important nucleus controlling the events
during the pre-collapse phase of iron cores of massive stars. The
calculated neutrino energy loss rates are around two orders of
magnitude enhanced as compared to large-scale shell model
calculations during the hydrostatic phases of stellar evolution.
This may affect the temperature, entropy and the lepton-to-baryon
ratio which becomes very important going into stellar collapse. We
will urge simulators to test run our reported weak interaction
rates presented here to check for some interesting outcome. We are
currently in a phase of extending the present work for other
nuclide of astrophysical importance and hope to report on the
outcome of these calculations in near future.

\textbf{ACKNOWLEDGMENTS}

This work is partially supported by the ICTP (Italy) through the
OEA-project-Prj-16.

\newpage
\textbf{Table (I):} The accuracy of the pn-QRPA model compared to
experimental data ($\beta^{+}$/EC, taken from Ref. [27]). $N$
denotes the number of experimentally known
 half-lives shorter than the limit in the second column, $n$ is the number (and percentage)
 of isotopes reproduced under the condition given in the first column, and $\bar{x}$ is the
 average deviation defined in the text.
 \begin{center}
\begin{tabular}{c|c|c|c|c|c}
Conditions & $T_{1/2}^{exp}(s) \leq$ & $N$ & $n$ & $n(\%)$ &
$\bar{x}$\\ \hline
$\forall x_{i} \leq$ 10 & 10$^{6}$ & 894 & 706 & 79.0 & 2.057 \\
                        & 60       & 327 & 304 & 93.0 & 1.718 \\
                        &  1       &  81 &  78 & 96.3 & 1.848 \\ \hline
$\forall x_{i} \leq$  2 & 10$^{6}$ & 894 & 489 & 54.7 & 1.363 \\
                        & 60       & 327 & 245 & 74.9 & 1.308 \\
                        &  1       &  81 &  59 & 72.8 & 1.230 \\
\end{tabular}
\end{center}
\newpage
\textbf{Table (II):} The accuracy of the pn-QRPA model compared to
experimental data ($\beta^{-}$, taken from Ref. [26]). $N$ denotes
the number of experimentally known
 half-lives shorter than the limit in the second column, $n$ is the number (and percentage)
  of isotopes reproduced under the condition given in the first column, and $\bar{x}$
  is the average deviation defined in the text.
\begin{center}
\begin{tabular}{c|c|c|c|c|c}
Conditions & $T_{1/2}^{exp}(s) \leq$ & $N$ & $n$ & $n(\%)$ &
$\bar{x}$\\ \hline
$\forall x_{i} \leq$ 10 & 10$^{6}$ & 654 & 472 & 72.2 & 1.85 $\pm$ 1.21 \\
                        & 60       & 325 & 313 & 96.3 & 1.67 $\pm$ 1.02 \\
                        &  1       & 106 & 105 & 99.1 & 1.44 $\pm$ 0.40 \\ \hline
$\forall x_{i} \leq$  5 & 10$^{6}$ & 654 & 456 & 69.7 & 1.68 $\pm$ 0.76 \\
                        & 60       & 325 & 307 & 94.5 & 1.56 $\pm$ 0.66 \\
                        &  1       & 106 & 105 & 99.1 & 1.44 $\pm$ 0.40 \\ \hline
$\forall x_{i} \leq$  3 & 10$^{6}$ & 654 & 420 & 64.2 & 1.50 $\pm$ 0.46 \\
                        & 60       & 325 & 295 & 90.8 & 1.46 $\pm$ 0.43 \\
                        &  1       & 106 & 105 & 99.1 & 1.44 $\pm$ 0.40 \\ \hline
$\forall x_{i} \leq$  2 & 10$^{6}$ & 654 & 369 & 56.4 & 1.37 $\pm$ 0.29 \\
                        & 60       & 325 & 267 & 82.2 & 1.36 $\pm$ 0.29 \\
                        &  1       & 106 &  96 & 90.6 & 1.35 $\pm$ 0.27 \\
\end{tabular}
\end{center}
\vspace{1.5in} \textbf{Table III:} Calculated excited states in
parent $^{55}Co$ using the pn-QRPA theory in units of MeV.
\begin{center}
\begin{tabular}{ccccccccccccccccccc} \\ \hline
0.0  &  & 2.17& & 2.57 & & 2.92 & &   3.08 & &  3.30 & & 3.87 & &
4.10& & 4.48 & &4.89\\
5.20  & & 5.47 & & 5.68  & & 5.99 & & 6.20 & & 6.50 & & 6.79 & &
7.08 & & 7.33 & &   7.65\\
7.88 & & 8.16 & &   8.42 & & 8.67 & & 8.96 & &  9.21 & & 9.48 & &
9.74& & 9.89 & & 10.00\\ \hline
\end{tabular}
\end{center}
\newpage
\textbf{Table IV:} Calculated B(GT$_{-}$) values from ground state
in $^{55}Co$. The energy scale refers to excitation energies in
daughter, $^{55}$Ni.
\begin{center}
\begin{tabular}{cc|cc|cc} \\ \hline
Energy(MeV)& B(GT$_{-}$) & Energy(MeV)& B(GT$_{-}$) &Energy(MeV)&
B(GT$_{-}$)
\\\hline
0.00&    3.06E-01&    6.36&    1.80E-02&   8.36&    5.37E-02\\
2.46&    1.40E+00&    6.66&    2.91E-02&   8.62&    2.45E-02\\
3.01&    4.48E-01&    6.85&    4.47E-02&   8.73&    3.78E-01\\
3.45&    1.04E-01&    7.00&    1.12E-02&   8.90&    3.26E-01\\
3.57&    1.80E-03&    7.14&    1.06E-02&   9.11&    2.94E+00\\
3.68&    1.79E-01&    7.43&    1.19E-01&   9.26&    1.59E+00\\
3.90&    1.99E-01&    7.59&    6.53E-03&   9.42&    1.00E+00\\
4.11&    2.25E-02&    7.76&    8.77E-02&   9.60&    2.11E+00\\
5.82&    1.93E-02&    7.92&    1.62E-02&   9.87&    4.84E-02\\
5.93&    3.89E-03&    8.04&    1.92E-01&   10.03&   7.67E-02\\
6.13&    1.58E-02&    8.25&    7.51E-02&                    \\
\hline
\end{tabular}
\end{center}
\newpage
\textbf{Table V:} Calculated positron capture, neutrino and
antineutrino energy loss rates on $^{55}Co$ for different selected
densities and temperatures in stellar matter. log($\rho Y_{e}$)
has units of $g/cm^{3}$, where $\rho$ is the baryon density and
$Y_{e}$ is the ratio of the lepton number to the baryon number.
Temperatures ($T_{9}$) are measured in $10^{9}$ K. $\lambda_{pc}$
are the positron capture rates $(sec^{-1})$. $\lambda_{\nu}$ are
the total neutrino energy loss rates $(MeV . s^{-1})$ due to
$\beta^{+}$ decay and electron capture. $\lambda_{\bar{\nu}}$ are
the total antineutrino energy loss rates $(MeV . s^{-1})$ due to
$\beta^{-}$ decay and positron capture. All calculated rates are
tabulated in logarithmic (to base 10) scale. In the table, -100
means that the rate is smaller than 10$^{-100}$.
\scriptsize{\begin{center}
\begin{tabular}{ccccc|ccccc|ccccc} \\ \hline
$log\rho Y_{e}$& $T_{9}$ & $\lambda_{pc}$ &
$\lambda_{\nu}$&$\lambda_{\bar{\nu}}$& $log\rho Y_{e}$& $T_{9}$ &
$\lambda_{pc}$ & $\lambda_{\nu}$ & $\lambda_{\bar{\nu}}$ &
$log\rho Y_{e}$& $T_{9}$ & $\lambda_{pc}$ &
$\lambda_{\nu}$&$\lambda_{\bar{\nu}}$  \\\hline
0.5& 0.5& -95.544& -3.375&  -95.515& 1 &  8.5& -5.531&  -0.308&  -5.082&  2&   4.5& -11.138& -2.004&  -10.894\\
0.5& 1  & -48.153& -3.376&  -48.102& 1 &  9  & -5.137&  -0.136&  -4.666&  2&   5 &  -10.008& -1.743&  -9.736\\
0.5& 1.5& -32.485& -3.366&  -32.41 & 1 & 9.5 &-4.776 & 0.03  &  -4.284&  2&   5.5& -9.067 & -1.501&  -8.768\\
0.5& 2  & -24.599& -3.305&  -24.498& 1 &  10&  -4.442&  0.191&   -3.93&   2&   6&   -8.269&  -1.274&  -7.942\\
0.5& 2.5& -19.828& -3.143&  -19.699& 1 &  15&  -2.053&  1.571&   -1.365&  2&   6.5& -7.579& -1.062&  -7.227\\
0.5& 3  & -16.614& -2.884&  -16.457& 1 &  20&  -0.553&  2.584&   0.264&   2&   7 &  -6.976&  -0.86&   -6.598\\
0.5& 3.5& -14.292& -2.585&  -14.107&1 &  25 & 0.506 &  3.342 &  1.425&   2 &  7.5& -6.441 & -0.668&  -6.039\\
0.5& 4  & -12.529& -2.287&  -12.315& 1 &  30&  1.301&   3.933&   2.304&   2&   8 &  -5.963&  -0.484&  -5.537\\
0.5& 4.5& -11.138& -2.006&  -10.896& 1.5& 0.5& -96.522& -3.375&  -96.493& 2&   8.5& -5.53 &  -0.307&  -5.081\\
0.5& 5  & -10.008& -1.745&  -9.738 & 1.5& 1  & -48.16&  -3.376&  -48.108& 2&   9 & -5.137 & -0.136&  -4.665\\
0.5& 5.5& -9.068 & -1.502&  -8.77  & 1.5& 1.5& -32.486& -3.366&  -32.41&  2&   9.5& -4.775& 0.03&    -4.283\\
0.5& 6  & -8.269 & -1.276&  -7.944&  1.5& 2  & -24.599& -3.305&  -24.497& 2&  10 & -4.442 & 0.191&  -3.929\\
0.5& 6.5& -7.58  & -1.063&  -7.229&  1.5& 2.5& -19.827& -3.142&  -19.698& 2&   15&  -2.052&  1.571&   -1.364\\
0.5& 7  & -6.977 & -0.862&  -6.601&  1.5& 3  & -16.614& -2.884&  -16.456& 2&   20&  -0.553& 2.585&   0.266\\
0.5& 7.5& -6.442&  -0.67 &  -6.042&  1.5& 3.5& -14.292& -2.584&  -14.106& 2&   25&  0.506 &  3.343&   1.427\\
0.5& 8  & -5.964&  -0.486&  -5.539&  1.5& 4  & -12.529& -2.286&  -12.314& 2&   30&  1.301 &  3.934&   2.306\\
0.5& 8.5& -5.531&  -0.309&  -5.084&  1.5& 4.5& -11.138& -2.004& -10.894& 2.5& 0.5& -97.522& -3.375& -97.493\\
0.5& 9  & -5.137&  -0.138&  -4.668&  1.5& 5  & -10.008& -1.743&  -9.736&  2.5& 1 &  -48.228& -3.375&  -48.177\\
0.5& 9.5& -4.776&  0.028 &  -4.286&  1.5& 5.5& -9.067&  -1.501&  -8.768&  2.5& 1.5& -32.49 &-3.366&  -32.414\\
0.5& 10 & -4.443&   0.19 &  -3.932&  1.5& 6  & -8.269&  -1.275&  -7.943&  2.5& 2&   -24.6  & -3.305&  -24.498\\
0.5& 15 & -2.054&   1.569&  -1.368&  1.5& 6.5& -7.579 & -1.062&  -7.227&  2.5&2.5& -19.828& -3.142&  -19.698\\
0.5& 20 & -0.554&   2.582&   0.261&   1.5& 7 &  -6.976&  -0.86&   -6.599&  2.5& 3&   -16.614& -2.883&  -16.456\\
0.5& 25 & 0.505 &   3.34 &   1.422&  1.5& 7.5& -6.441 &-0.668 & -6.039 &2.5& 3.5& -14.292& -2.584&  -14.106\\
0.5& 30 & 1.3   &   3.931&   2.301&   1.5& 8 &  -5.963&  -0.484&  -5.537&  2.5& 4&   -12.529& -2.285&  -12.314\\
1  & 0.5& -96.024& -3.375&  -95.995& 1.5& 8.5& -5.53 &  -0.307 & -5.082&  2.5& 4.5& -11.138& -2.004&  -10.894\\
1  & 1  & -48.155& -3.376&  -48.103& 1.5& 9 &  -5.137&  -0.136&  -4.666& 2.5& 5 &  -10.008& -1.743&  -9.736\\
1  & 1.5& -32.485& -3.366&  -32.409& 1.5& 9.5& -4.775&  0.03  &  -4.283&  2.5& 5.5& -9.067&  -1.501& -8.768\\
1  & 2  & -24.599& -3.305&  -24.497& 1.5& 10 & -4.442&  0.191&   -3.929& 2.5& 6  & -8.269 & -1.274 & -7.942\\
1  & 2.5& -19.827& -3.142&  -19.698& 1.5& 15 & -2.053&  1.571&   -1.365&  2.5& 6.5& -7.579&  -1.062&  -7.227\\
1  & 3  & -16.614& -2.884&  -16.456& 1.5& 20 & -0.553&  2.585&   0.265&   2.5& 7 &  -6.976&  -0.86 &  -6.598\\
1  & 3.5& -14.292& -2.584&  -14.106& 1.5& 25 & 0.506 &  3.343&   1.426&   2.5& 7.5& -6.441&  -0.668&  -6.039\\
1  & 4  & -12.529& -2.286&  -12.314& 1.5& 30 & 1.301&   3.934&  2.305 &  2.5& 8&   -5.963&  -0.484&  -5.537\\
1  & 4.5& -11.138& -2.005&  -10.894& 2&  0.5& -97.022& -3.375&  -96.993& 2.5& 8.5& -5.53 &  -0.307&  -5.081\\
1  & 5  & -10.008& -1.744&  -9.736 & 2&   1 &  -48.176& -3.376&  -48.125& 2.5& 9&  -5.137&  -0.136&  -4.665\\
1  & 5.5& -9.067 & -1.501&  -8.768 & 2&   1.5& -32.487&-3.366&  -32.411& 2.5& 9.5& -4.775&  0.03&    -4.283\\
1  & 6  & -8.269&  -1.275&  -7.943 & 2&   2  & -24.599& -3.305&  -24.497&2.5& 10&  -4.442& 0.192&   -3.929\\
1  & 6.5& -7.58 &  -1.062&  -7.228 & 2&   2.5& -19.827& -3.142&  -19.698& 2.5& 15& -2.052&  1.572&   -1.364\\
1  & 7  & -6.976&  -0.861&  -6.599 & 2&   3  & -16.614&-2.884 & -16.456& 2.5& 20&  -0.553&  2.585&   0.266\\
1  & 7.5& -6.442&  -0.669&  -6.04  & 2&   3.5& -14.292& -2.584&  -14.106& 2.5& 25&  0.506&   3.343&   1.427\\
1  & 8  & -5.963&  -0.485&  -5.538 & 2&   4  & -12.529& -2.286& -12.313& 2.5& 30&  1.302&   3.935&   2.306\\
\end{tabular}
\end{center}
\newpage
\begin{center}
\begin{tabular}{ccccc|ccccc|ccccc} \\ \hline
$log\rho Y_{e}$& $T_{9}$ & $\lambda_{pc}$ &
$\lambda_{\nu}$&$\lambda_{\bar{\nu}}$& $log\rho Y_{e}$& $T_{9}$ &
$\lambda_{pc}$ & $\lambda_{\nu}$ & $\lambda_{\bar{\nu}}$ &
$log\rho Y_{e}$& $T_{9}$ & $\lambda_{pc}$ &
$\lambda_{\nu}$&$\lambda_{\bar{\nu}}$  \\\hline
3&   0.5&  -98.023&  -3.374 &  -97.994&  4&    0.5&  -99.036&  -3.364&   -99.006& 5&    0.5&  -100&     -3.281&   -100\\
3&    1 &   -48.382&  -3.375&   -48.331&  4&    1 &   -49.204&  -3.366&   -49.153&  5&    1&    -50.236&  -3.285&  -50.185\\
3&    1.5&  -32.501&  -3.366&   -32.425&  4&    1.5&  -32.643&  -3.36 &   -32.567&  5&    1.5&  -33.377&  -3.276&   -33.301\\
3&    2  &  -24.603&  -3.305&   -24.501&  4&    2  &  -24.634&  -3.299&   -24.532&  5&    2&    -24.923&  -3.229&   -24.821\\
3&    2.5&  -19.828&  -3.142&  -19.699&  4&    2.5&  -19.84 &  -3.137 & -19.71&   5&    2.5&  -19.953&  -3.084 &  -19.823\\
3&    3  &  -16.614&  -2.883&   -16.457&  4&    3 &   -16.62&   -2.88&    -16.462&  5&    3&    -16.67& 3 -2.843&   -16.515\\
3&    3.5&  -14.292&  -2.584&  -14.106&  4&    3.5&  -14.295& -2.581&   -14.109&  5 &   3.5&  -14.325&  -2.558 &  -14.138\\
3&    4  &  -12.529&  -2.285&   -12.314&  4&    4 &   -12.53&   -2.284&   -12.315&  5&    4&    -12.549&  -2.269&   -12.333\\
3&    4.5&  -11.138&  -2.004&   -10.894&  4&    4.5&  -11.139&  -2.003&   -10.895&  5&    4.5&  -11.151&  -1.993&   -10.907\\
3&    5  &  -10.008&  -1.743&   -9.736&   4&    5 &   -10.008&  -1.742&   -9.737&   5&    5&   -10.017&  -1.735&   -9.745\\
3&    5.5&  -9.067 &  -1.501&  -8.768 &  4&    5.5&  -9.068&  -1.5   &  -8.768&   5 &   5.5&  -9.074&   -1.495&   -8.774\\
3&    6  &  -8.269&   -1.274&  -7.942 &  4&    6   & -8.269&   -1.274&   -7.943&   5&    6 &   -8.274&   -1.27 &   -7.947\\
3&    6.5& -7.579 &  -1.061&   -7.227&   4&    6.5&  -7.58&    -1.061&   -7.227&   5&    6.5&  -7.583&   -1.058&   -7.231\\
3&    7  &  -6.976&   -0.86 &   -6.598&   4&    7 &   -6.976&   -0.86&    -6.598&   5&    7&    -6.979&   -0.857&   -6.601\\
3&    7.5&  -6.441&   -0.668&  -6.039 &  4&    7.5&  -6.442&  -0.668 &  -6.039&   5 &   7.5&  -6.444&   -0.666&   -6.041\\
3&    8 &   -5.963&   -0.484&   -5.537&   4&    8 &   -5.963&   -0.484&  -5.537&   5 &  8 &   -5.965&   -0.482&   -5.539\\
3&    8.5&  -5.53 &   -0.307&  -5.081&   4&    8.5&  -5.53&    -0.307&   -5.081&  5 &   8.5&  -5.532&   -0.305&   -5.083\\
3&    9  &  -5.136&   -0.136&   -4.665&   4&    9 &   -5.137&   -0.135&   -4.665&   5&    9&   -5.138&   -0.134&   -4.666\\
3&    9.5&  -4.775&  0.03  &   -4.283&   4&    9.5&  -4.775&  0.031 &   -4.283&   5 &   9.5&  -4.776&   0.032&    -4.284\\
3&    10&   -4.442&   0.192&   -3.929&   4&    10 &  -4.442&   0.192&    -3.929&  5&    10&   -4.443&   0.193&    -3.929\\
3&    15&   -2.052&   1.572&   -1.364&   4&    15 &  -2.052&   1.572 &   -1.364&   5&    15&   -2.053&   1.572&    -1.364\\
3&    20&   -0.552&   2.585&    0.266&   4&    20&   -0.552&   2.585&    0.266&    5&    20&   -0.553&   2.586&    0.266\\
3&    25&   0.506 &   3.343&    1.427&    4&    25&   0.507&    3.343&    1.427 &   5&    25&   0.506&    3.343&    1.427\\
3&    30&   1.302 &   3.935&    2.306&    4&    30&  1.302&    3.935 &   2.306&    5 &   30 &  1.302&    3.935&    2.306\\

3.5&  0.5&  -98.526&  -3.372&   -98.497&  4.5&  0.5&  -99.565&  -3.342&   -99.536&  5.5&  0.5&  -100&     -3.144&   -100\\
3.5&  1 &   -48.73 &  -3.373&   -48.679& 4.5& 1  &  -49.71&   -3.345 &  -49.659&  5.5&  1&    -50.815&  -3.14 &   -50.764\\
3.5&  1.5&  -32.536&  -3.365&   -32.46&   4.5&  1.5&  -32.921&  -3.341&   -32.845&  5.5&  1.5&  -33.907&  -3.111&   -33.831\\
3.5&  2 &   -24.61 &  -3.303&   -24.508&  4.5&  2 &   -24.709&  -3.284&  -24.607&  5.5&  2&    -25.344&  -3.059&   -25.242\\
3.5&  2.5&  -19.831&  -3.141&  -19.702&  4.5&  2.5&  -19.867&  -3.125 &  -19.738& 5.5&  2.5&  -20.191&  -2.951&   -20.061\\
3.5&  3 &   -16.615&  -2.882&   -16.458&  4.5&  3 &   -16.632&  -2.871&  -16.475& 5.5&  3 &   -16.798&  -2.755&   -16.64\\
3.5&  3.5&  -14.293&  -2.583&  -14.107&  4.5&  3.5&  -14.302&  -2.576 &  -14.116&  5.5&  3.5&  -14.395&  -2.502&   -14.209\\
3.5&  4  &  -12.529&  -2.285&   -12.314&  4.5&  4&    -12.535&  -2.28 &   -12.32&   5.5&  4&    -12.592&  -2.232&   -12.377\\
3.5&  4.5&  -11.138&  -2.004&   -10.894& 4.5&  4.5&  -11.142&  -2.001 &  -10.898&  5.5&  4.5&  -11.18&   -1.968&   -10.936\\
3.5&  5 &   -10.008&  -1.743&   -9.736&   4.5&  5 &   -10.01&   -1.741&   -9.739&   5.5&  5&    -10.037&  -1.718&   -9.765\\
3.5&  5.5&  -9.067&   -1.501&   -8.768&   4.5&  5.5& -9.069&   -1.499&   -8.77&    5.5&  5.5&  -9.088&   -1.482&   -8.789\\
3.5&  6  &  -8.269 &  -1.274&  -7.942&   4.5&  6  &  -8.27&    -1.273&   -7.944&   5.5&  6 &   -8.284&   -1.26&    -7.958\\
3.5&  6.5&  -7.579&   -1.061&   -7.227&   4.5&  6.5&  -7.581&   -1.06&   -7.228&   5.5&  6.5&  -7.592&   -1.051&   -7.239 \\
3.5&  7  &  -6.976&   -0.86 &   -6.598&   4.5&  7 &   -6.977&   -0.859&   -6.599&   5.5&  7&    -6.986&   -0.851&   -6.608\\
3.5&  7.5&  -6.441&   -0.668&   -6.039&   4.5&  7.5&  -6.442&   -0.667&   -6.04&    5.5&  7.5&  -6.449&   -0.661&   -6.047\\
3.5&  8 &   -5.963&   -0.484&   -5.537&   4.5&  8 &   -5.963&   -0.484&   -5.537&  5.5&  8 &   -5.969&   -0.478&   -5.543\\
3.5&  8.5&  -5.53 &   -0.307&   -5.081&   4.5&  8.5&  -5.531&   -0.306&   -5.082&   5.5&  8.5&  -5.535&   -0.302&   -5.086\\
3.5&  9  &  -5.137&   -0.136&   -4.665&   4.5&  9 &   -5.137&   -0.135&   -4.666&   5.5&  9&    -5.141&   -0.132&   -4.669\\
3.5&  9.5&  -4.775&  0.03 &    -4.283 &  4.5&  9.5&  -4.776&   0.031&   -4.283 &  5.5&  9.5&  -4.779&   0.034 &   -4.286\\
3.5&  10 &  -4.442&   0.192&    -3.929&   4.5&  10&   -4.442&   0.192&    -3.929&   5.5&  10&   -4.445&   0.195&    -3.932\\
3.5&  15&   -2.052&   1.572&    -1.364&  4.5 & 15 &  -2.052&   1.572 &   -1.364 &  5.5&  15&   -2.053&   1.573 &   -1.365\\
3.5&  20&   -0.552&   2.585&    0.266 &   4.5&  20&   -0.552&   2.585&    0.266&    5.5&  20&   -0.553&   2.586&    0.266\\
3.5&  25&   0.506 &   3.343&    1.427&    4.5&  25&   0.507&    3.343&    1.427&    5.5&  25&   0.506&    3.343&   1.427\\
3.5&  30&   1.302 &   3.935&    2.306&    4.5&  30&  1.302&    3.935 &   2.306&    5.5&  30&   1.302 &   3.935&    2.306\\
\end{tabular}
\end{center}
\newpage

\begin{center}
\begin{tabular}{ccccc|ccccc|ccccc} \\ \hline
$log\rho Y_{e}$& $T_{9}$ & $\lambda_{pc}$ &
$\lambda_{\nu}$&$\lambda_{\bar{\nu}}$& $log\rho Y_{e}$& $T_{9}$ &
$\lambda_{pc}$ & $\lambda_{\nu}$ & $\lambda_{\bar{\nu}}$ &
$log\rho Y_{e}$& $T_{9}$ & $\lambda_{pc}$ &
$\lambda_{\nu}$&$\lambda_{\bar{\nu}}$  \\\hline
6&   0.5& -100&    -2.891&  -100&    7&   0.5& -100&    -1.928&  -100&    8&   0.5& -100&    -0.379&  -100\\
6&   1 &  -51.539& -2.864&  -51.488& 7&   1 &  -54.202& -1.879&  -54.151& 8&   1 &  -60.433& -0.338&  -60.382\\
6&   1.5& -34.514& -2.805&  -34.438& 7&   1.5& -36.425& -1.79&   -36.349& 8&   1.5& -40.629& -0.265&  -40.553\\
6&   2  & -25.89&  -2.729&  -25.788&7 &  2  & -27.455& -1.695&  -27.353& 8&   2 &  -30.662& -0.195&  -30.56\\
6&   2.5& -20.643& -2.636&  -20.514& 7&   2.5& -22.01& -1.602&  -21.881& 8&   2.5& -24.631& -0.134&  -24.502\\
6&   3 &  -17.116& -2.507&  -16.958&7&   3  & -18.33& -1.51 &  -18.172& 8&   3  & -20.57&  -0.078&  -20.412\\
6&   3.5& -14.6 &  -2.331&  -14.414& 7&   3.5& -15.66&  -1.416&  -15.474& 8&   3.5& -17.635& -0.024&  -17.449\\
6&   4 &  -12.725& -2.117& -12.51 & 7&   4  & -13.625& -1.319&  -13.41&  8&   4&   -15.406& 0.03&    -15.19\\
6&   4.5& -11.269& -1.889&  -11.025& 7&   4.5& -12.017& -1.217&  -11.773& 8&   4.5& -13.647& 0.086&   -13.403\\
6&   5  & -10.1  & -1.662&  -9.828&  7&   5 &  -10.711& -1.108&  -10.439& 8&   5&   -12.217& 0.144&   -11.945\\
6&   5.5& -9.134&  -1.441&  -8.834&  7&   5.5& -9.629&  -0.992&  -9.329&  8&   5.5& -11.027& 0.204&   -10.727\\
6&   6 &  -8.319&  -1.23&   -7.992& 7&   6  & -8.717&  -0.867 & -8.391& 8&   6 &  -10.016& 0.267&   -9.69\\
6&   6.5& -7.618&  -1.027& -7.266&  7&   6.5& -7.94&   -0.735&  -7.587&  8&   6.5& -9.145&  0.333&   -8.792\\
6&   7 &  -7.007&  -0.833&  -6.629&  7&   7 &  -7.267&  -0.596&  -6.889&  8 &  7&   -8.382&  0.402&   -8.004\\
6&   7.5& -6.466&  -0.646& -6.063 & 7&   7.5& -6.679&  -0.452&  -6.276&  8&   7.5& -7.706&  0.475&   -7.304\\
6&   8 &  -5.983&  -0.466&  -5.557&  7&   8 &  -6.158&  -0.306&  -5.732&  8&   8 &  -7.102&  0.551&   -6.676\\
6&  8.5& -5.547&  -0.292&  -5.098 & 7&   8.5& -5.693&  -0.158&  -5.244&  8&   8.5& -6.557&  0.63&    -6.108\\
6&   9 &  -5.15&   -0.123&  -4.679&  7&   9&   -5.273&  -0.01&   -4.802&  8&   9&   -6.062&  0.713&   -5.591\\
6&   9.5& -4.787&  0.041&   -4.294&  7&   9.5& -4.891&  0.137&   -4.398& 8&   9.5& -5.61 &  0.798&   -5.117\\
6&   10&  -4.452&  0.201&   -3.938&  7&   10 & -4.541&  0.283&   -4.027&  8&   10&  -5.195&  0.887&   -4.681\\
6&   15&  -2.055&  1.575&   -1.367&  7&   15 & -2.081&  1.599&   -1.392&  8&   15&  -2.327&  1.834&   -1.638\\
6&   20&  -0.554&  2.587&   0.265 &  7&   20 & -0.564&  2.597&   0.254&   8&   20&  -0.671&  2.7& 0.148\\
6&   25&  0.506 &  3.344&   1.427 &  7&   25 & 0.5& 3.349 &  1.421&        8&   25&  0.446&   3.403&   1.367\\
6&   30&  1.301 &  3.935&   2.306 &  7&   30 & 1.298&   3.938&   2.303&   8&   30&  1.267 & 3.969&   2.272\\

6.5& 0.5& -100&    -2.489&  -100 &   7.5& 0.5& -100&    -1.22&   -100&    8.5& 0.5& -100  &  0.554&   -100\\
6.5& 1 &  -52.585& -2.444&  -52.534& 7.5& 1  & -56.687& -1.173&  -56.636& 8.5& 1&   -66.005& 0.587&   -65.954\\
6.5& 1.5& -35.297& -2.362&  -35.221& 7.5& 1.5& -38.112& -1.088&  -38.036& 8.5& 1.5& -44.358& 0.647&   -44.282\\
6.5& 2 &  -26.557&-2.268&  -26.455& 7.5& 2   &-28.753& -1.003&  -28.651& 8.5& 2 &  -33.473& 0.703&   -33.371\\
6.5& 2.5& -21.24 & -2.171&  -21.111& 7.5& 2.5& -23.082& -0.924&  -22.952& 8.5& 2.5& -26.895& 0.749&   -26.765\\
6.5& 3 &  -17.642& -2.067& -17.484& 7.5& 3  & -19.256& -0.849 & -19.099& 8.5& 3 &  -22.471& 0.789 &  -22.313\\
6.5& 3.5& -15.036& -1.951&  -14.85&  7.5& 3.5& -16.487& -0.775&  -16.3 &  8.5& 3.5& -19.28&  0.826&   -19.094\\
6.5& 4  & -13.063& -1.817&  -12.847& 7.5& 4  & -14.378& -0.699&  -14.163& 8.5& 4&   -16.86&  0.863&   -16.645\\
6.5& 4.5& -11.521& -1.664&  -11.278& 7.5& 4.5& -12.712& -0.622&  -12.468& 8.5& 4.5& -14.955& 0.9& -14.711\\
6.5& 5 & -10.287& -1.494&  -10.015& 7.5& 5   &-11.354& -0.542 & -11.083& 8.5& 5&   -13.41&  0.939&   -13.138\\
6.5& 5.5& -9.274&  -1.316&  -8.974&  7.5& 5.5& -10.223& -0.459&  -9.924&  8.5& 5.5& -12.127& 0.979&   -11.827\\
6.5& 6 &  -8.426&  -1.134&  -8.099&  7.5& 6 &  -9.262&  -0.374& -8.936&  8.5& 6&   -11.04 & 1.023&   -10.713\\
6.5& 6.5& -7.701&  -0.952&  -7.348&  7.5& 6.5& -8.433&  -0.284&  -8.08&  8.5& 6.5& -10.104& 1.069&  -9.751\\
6.5& 7 &  -7.072&  -0.773&  -6.694& 7.5& 7  & -7.709&  -0.191 & -7.331&  8.5& 7&   -9.287&  1.119&   -8.909\\
6.5& 7.5& -6.519&  -0.598&  -6.116&  7.5& 7.5& -7.071&  -0.092& -6.668&  8.5& 7.5& -8.564&  1.173&   -8.162\\
6.5& 8 &  -6.026&  -0.427&  -5.6  &  7.5&8  & -6.503&  0.011 &  -6.077&  8.5& 8&   -7.919&  1.23&    -7.493\\
6.5& 8.5& -5.582&  -0.259&  -5.133&  7.5& 8.5& -5.995&  0.12&    -5.545&  8.5& 8.5& -7.338&  1.29&    -6.889\\
6.5& 9 &  -5.18 &  -0.096&  -4.709&  7.5& 9  & -5.536&  0.233&   -5.065&  8.5& 9 &  -6.811&  1.355&   -6.34\\
6.5& 9.5& -4.812&  0.064&   -4.319&  7.5&9.5& -5.12&   0.349 &  -4.628&  8.5& 9.5& -6.329 & 1.422&   -5.836\\
6.5& 10&  -4.473&  0.221&   -3.96 &  7.5& 10&  -4.741&  0.469&   -4.227&  8.5& 10&  -5.885&  1.492&   -5.371\\
6.5& 15&  -2.061&  1.58 &   -1.373&  7.5& 15&  -2.142&  1.658&   -1.454& 8.5& 15&  -2.761&  2.239 &  -2.072\\
6.5& 20&  -0.556&  2.589&   0.263&   7.5& 20&  -0.59&   2.622&   0.229&   8.5& 20&  -0.906&  2.926&   -0.087\\
6.5& 25&  0.505 &  3.345&   1.426 &  7.5& 25&  0.487&   3.362&   1.408&  8.5& 25&  0.318 &  3.527&   1.239\\
6.5& 30&  1.301 &  3.936&   2.305&   7.5& 30&  1.291&   3.946&   2.296&   8.5& 30&  1.192&   4.043&   2.196\\

\end{tabular}
\end{center}
\newpage
\begin{center}
\begin{tabular}{ccccc|ccccc|ccccc} \\ \hline
$log\rho Y_{e}$& $T_{9}$ & $\lambda_{pc}$ &
$\lambda_{\nu}$&$\lambda_{\bar{\nu}}$& $log\rho Y_{e}$& $T_{9}$ &
$\lambda_{pc}$ & $\lambda_{\nu}$ & $\lambda_{\bar{\nu}}$ &
$log\rho Y_{e}$& $T_{9}$ & $\lambda_{pc}$ &
$\lambda_{\nu}$&$\lambda_{\bar{\nu}}$  \\\hline
9&   0.5& -100&    1.546&   -100&    9.5& 8.5& -9.889&  2.978&   -9.44&   10.5&    4.5& -29.371& 4.964&   -29.127\\
9&   1  & -74.238& 1.573&  -74.187& 9.5& 9 &  -9.237&  3.01&    -8.766&  10.5&    5 &  -26.41&  4.97 &   -26.139\\
9&   1.5& -49.856& 1.623&   -49.78&  9.5& 9.5& -8.644&  3.044&   -8.151&  10.5&    5.5& -23.972& 4.975&   -23.672\\
9&   2  & -37.606& 1.668&   -37.504& 9.5& 10&  -8.102&  3.082&   -7.588&  10.5&    6 &  -21.924& 4.98&    -21.597\\
9&   2.5& -30.211& 1.704&   -30.082& 9.5& 15&  -4.357&  3.568&   -3.668&  10.5&    6.5& -20.177& 4.986&   -19.824\\
9&   3 & -25.245 &1.734&   -25.087& 9.5& 20 & -2.141&  4.038&   -1.323&  10.5&    7  & -18.666& 4.993 &  -18.289\\
9&   3.5& -21.668& 1.76&    -21.482& 9.5& 25&  -0.627&  4.412&    0.294&  10.5&    7.5& -17.345& 5 &      -16.943\\
9&   4 &  -18.96 & 1.785&   -18.745& 9.5& 30 &  0.486&  4.718&    1.491&  10.5&    8 &  -16.178& 5.009 &  -15.752\\
9&   4.5& -16.832& 1.81&    -16.589& 10&  0.5& -100&    3.739&   -100&    10.5&    8.5&-15.137& 5.02&    -14.688\\
9&   5 &  -15.11 & 1.835&   -14.838& 10&  1  & -100&    3.756&   -100&    10.5&    9 &  -14.202& 5.033&   -13.731\\
9&   5.5& -13.683& 1.862&   -13.383& 10&  1.5& -69.826& 3.787&   -69.75&  10.5&    9.5& -13.356& 5.047&   -12.864\\
9&   6  & -12.477& 1.89&    -12.15&  10&  2 &  -52.595& 3.815&   -52.493& 10.5&    10 & -12.587& 5.064&   -12.073\\
9&   6.5& -11.441& 1.92&    -11.088& 10 & 2.5& -42.214& 3.836&   -42.085& 10.5&    15&  -7.419&  5.341&   -6.73\\
9&  7  & -10.539& 1.954&  -10.161& 10&  3&   -35.259& 3.852 &  -35.101& 10.5 &   20&  -4.511&  5.662&   -3.693\\
9&   7.5& -9.744&  1.99&    -9.341&  10& 3.5& -30.263& 3.865&   -30.077& 10.5&    25&  -2.592&  5.92&    -1.672\\
9&   8 &  -9.035&  2.03&    -8.609&  10&  4 &  -26.493& 3.875&   -26.278& 10.5&    30&  -1.212&  6.123&   -0.207\\
9&   8.5& -8.399& 2.073&   -7.95 &  10&  4.5& -23.54&  3.885 &  -23.297& 11&  0.5& -100&    5.914&   -100\\
9&   9 &  -7.823&  2.12&    -7.351&  10&  5 &  -21.159& 3.894&   -20.887& 11&  1&  -100&    5.927&   -100\\
9&   9.5& -7.297&  2.171&   -6.804&  10&  5.5& -19.194& 3.904&   -18.895& 11&  1.5& -100&    5.95&    -100\\
9&   10 &-6.815&  2.224 & -6.301&  10 & 6&   -17.541& 3.913 &  -17.215& 11&  2 &  -84.902& 5.971&   -84.8\\
9&   15 & -3.44 &  2.841&   -2.752&  10&  6.5& -16.128& 3.923&   -15.776& 11&  2.5& -68.065& 5.986&   -67.936\\
9&   20 & -1.406&  3.395&   -0.588&  10&  7 &  -14.904& 3.935&   -14.526& 11&  3&  -56.807& 5.996&   -56.649\\
9&   25 & -0.021&  3.853&   0.899&   10&  7.5& -13.83&  3.948&   -13.427& 11&  3.5& -48.739& 6.004&   -48.552\\
9&   30 & 0.971&   4.258&   1.975&   10&  8 &  -12.879& 3.962&   -12.452& 11&  4&  -42.665& 6.01&    -42.449\\
9.5& 0.5& -100 &   2.617&   -100 &   10&  8.5& -12.028& 3.979&   -11.579& 11&  4.5& -37.92&  6.015&   -37.677\\
9.5& 1 &  -86.36&  2.638&   -86.308& 10& 9  & -11.263& 3.999&   -10.792& 11&  5 &  -34.107& 6.019&   -33.835\\
9.5& 1.5& -57.943& 2.678&   -57.867& 10&  9.5& -10.568& 4.021&   -10.076& 11&  5.5& -30.971& 6.022&   -30.671\\
9.5& 2 &  -43.678&2.714 &  -43.576& 10&  10 & -9.935&  4.046 &  -9.421 & 11&  6 &  -28.342& 6.026&   -28.016\\
9.5& 2.5& -35.076& 2.742&   -34.947& 10&  15&  -5.622&  4.411&   -4.933&  11&  6.5& -26.104& 6.03&    -25.751\\
9.5& 3 &  -29.306& 2.764&  -29.148& 10&  20 & -3.133& 4.8 &    -2.314&  11&  7 &  -24.172& 6.034&   -23.794\\
9.5& 3.5& -25.156& 2.782&   -24.97&  10&  25&  -1.459&  5.112&   -0.538&  11&  7.5& -22.486& 6.039&   -22.084\\
9.5& 4 & -22.019& 2.799&   -21.804& 10 & 30 & -0.237&  5.362 &   0.768 &  11&  8 &  -21& 6.045&      -20.574\\
9.5& 4.5& -19.558& 2.815&   -19.315& 10.5& 0.5& -100&    4.848&   -100 &   11&  8.5& -19.678& 6.053&   -19.229\\
9.5& 5 &  -17.571& 2.831&   -17.299& 10.5& 1 &  -100&    4.862&   -100&    11& 9&   -18.493& 6.062&   -18.022\\
9.5& 5.5& -15.927& 2.847&   -15.628& 10.5& 1.5& -87.276& 4.888&   -87.2 &  11&  9.5& -17.424& 6.073&   -16.931\\
9.5& 6  & -14.541& 2.864&   -14.215& 10.5& 2 &  -65.686& 4.911&   -65.584& 11&  10&  -16.453& 6.085&   -15.94\\
9.5& 6.5& -13.354& 2.882&   -13.002& 10.5& 2.5& -52.69 & 4.928&   -52.56&  11&  15&  -10.017& 6.311&   -9.328\\
9.5& 7 &  -12.323& 2.903&   -11.945& 10.5& 3  & -43.992& 4.94&    -43.834& 11&  20&  -6.48&   6.585&   -5.662\\
9.5& 7.5& -11.416& 2.925&   -11.013& 10.5& 3.5& -37.752&4.95 &   -37.566&11&  25& -4.189&  6.805&   -3.269\\
9.5& 8&   -10.61&  2.95&    -10.184& 10.5& 4 &  -33.049& 4.957&
-32.834& 11&  30&  -2.564&  6.971&   -1.56\\\hline
\end{tabular}
\end{center}}

\begin{figure}[htbp]
\caption{Positron capture rates on $^{55}$Co, as a function of
stellar temperatures, for different selected densities . Densities
are in units of $gcm^{-3}$. Temperatures are measured in $10^{9}$
K and log$_{10}\lambda_{pc}$ represents the log of positron
capture rates in units of $sec^{-1}$.}
\begin{center}
\begin{tabular}{c}
\includegraphics[width=0.8\textwidth]{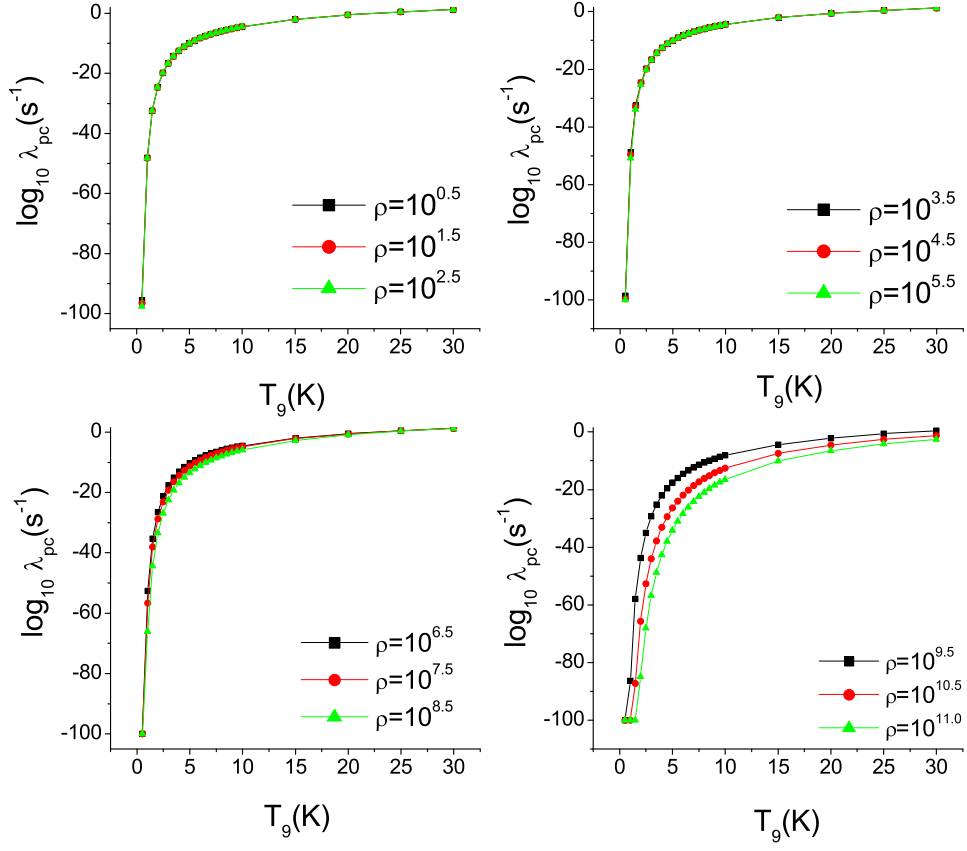}
\end{tabular}
\end{center}
\end{figure}

\begin{figure}[htbp]
\caption{Neutrino energy loss rates on $^{55}$Co, as a function of
stellar temperatures, for different selected densities . Densities
are in units of $gcm^{-3}$. Temperatures are measured in $10^{9}$
K and log$_{10}\lambda_{\nu}$ represents the log of neutrino
energy loss rates in units of $MeV.sec^{-1}$.}
\begin{center}
\begin{tabular}{c}
\includegraphics[width=0.8\textwidth]{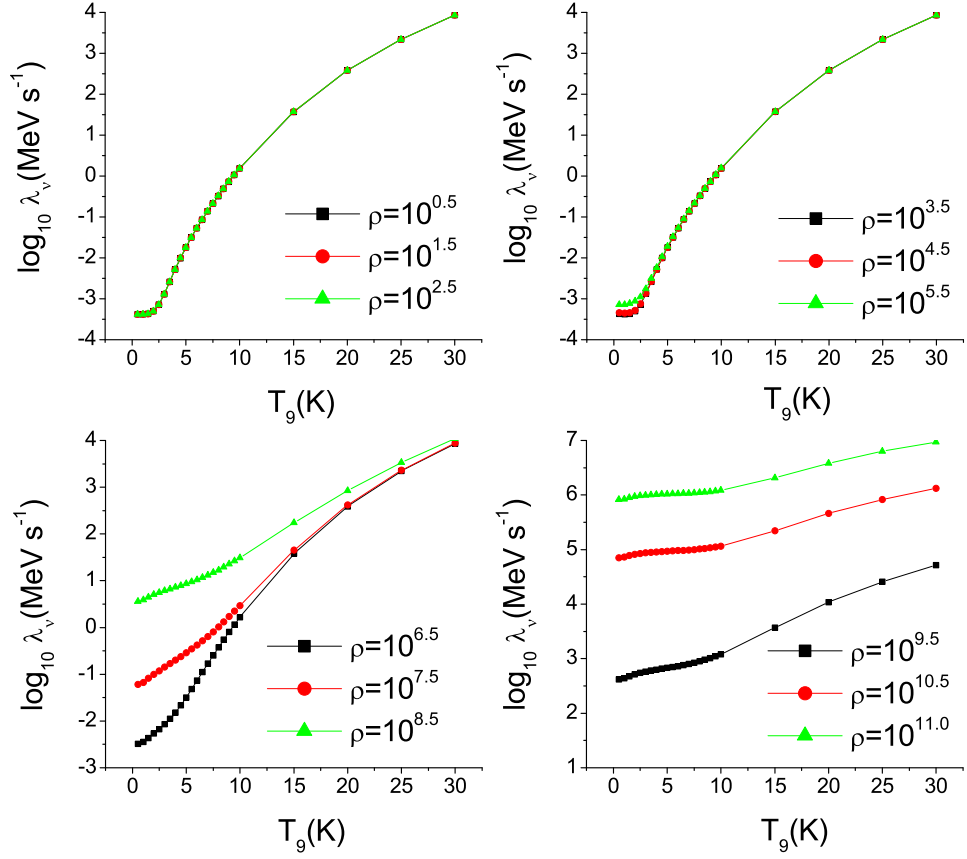}
\end{tabular}
\end{center}
\end{figure}
\newpage
\begin{figure}[htbp]
\caption{same as Fig.2 but for antineutrino energy loss rates.}
\begin{center}
\begin{tabular}{c}
\includegraphics[width=0.8\textwidth]{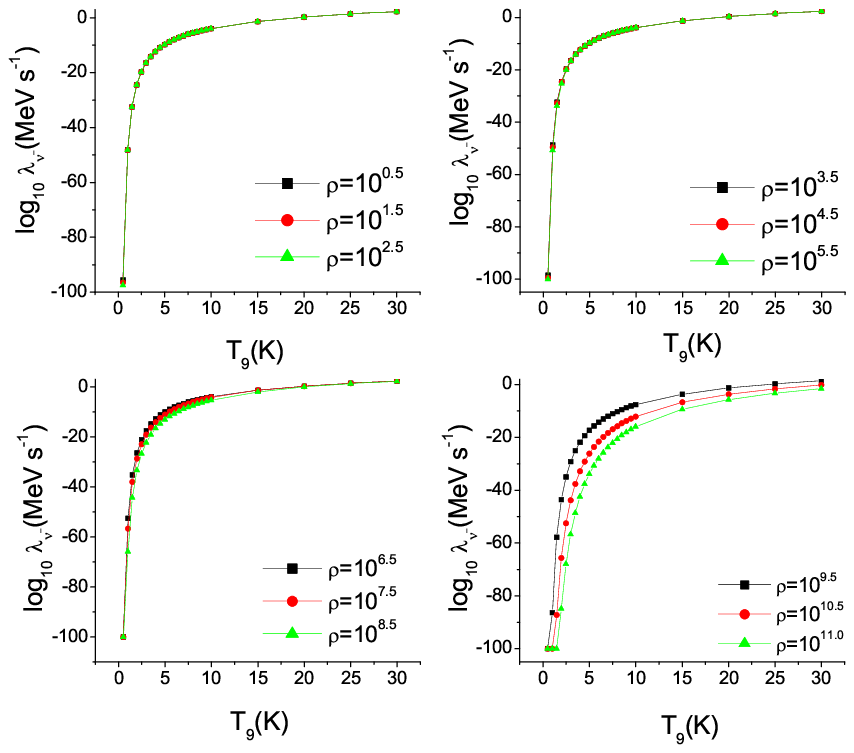}
\end{tabular}
\end{center}
\end{figure}

\begin{figure}[htbp]
\caption{Comparison of neutrino energy loss rates with those of
large-scale shell model [12] and FFN [10] calculations as a
function of stellar temperatures and densities.}
\begin{center}
\begin{tabular}{c}
\includegraphics[width=1.0\textwidth]{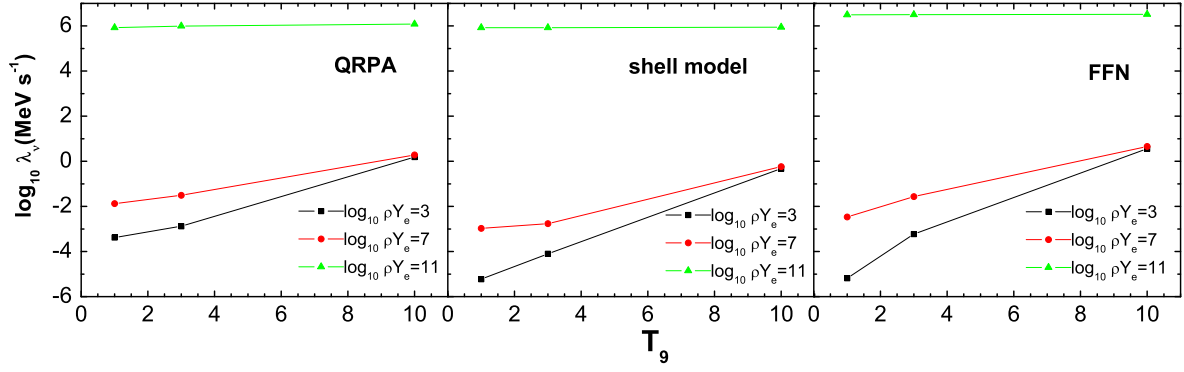}
\end{tabular}
\end{center}
\end{figure}

\begin{figure}[htbp]
\caption{Same as Fig.4 but for antineutrino energy loss rates.}
\begin{center}
\begin{tabular}{c}
\includegraphics[width=1.0\textwidth]{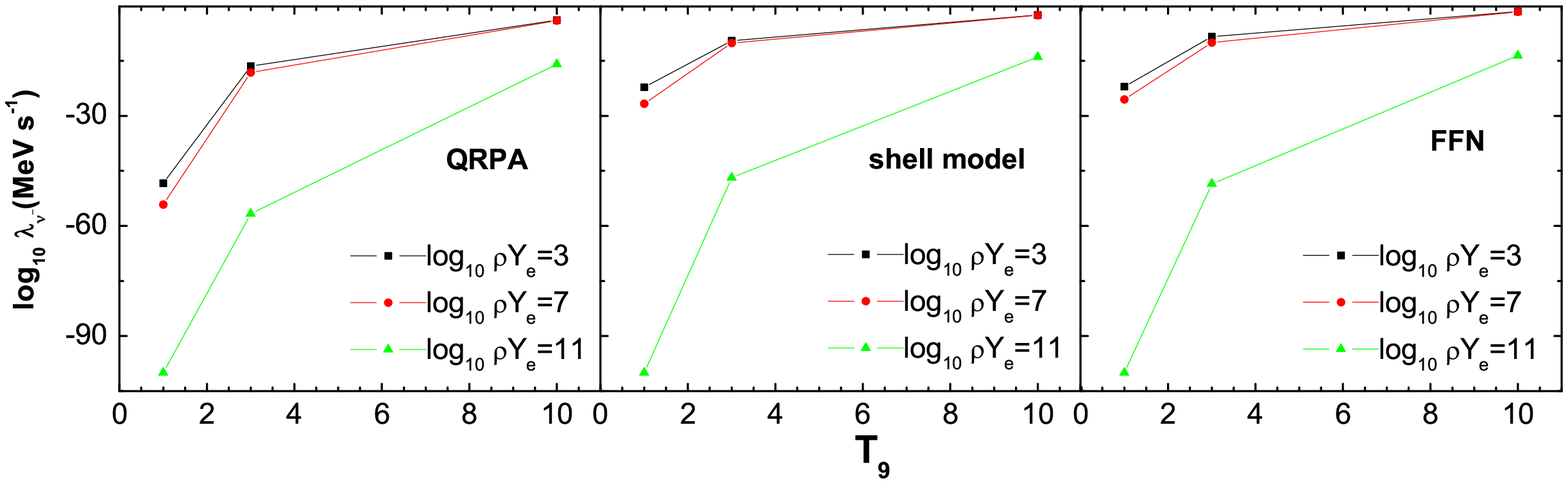}
\end{tabular}
\end{center}
\end{figure}

\begin{figure}[htbp]
\caption{Cumulative sum of the B(GT$_{-}$) values. The energy
scale refers to excitation energies in daughter $^{55}$Ni. }
\begin{center}
\begin{tabular}{c}
\includegraphics[width=0.8\textwidth]{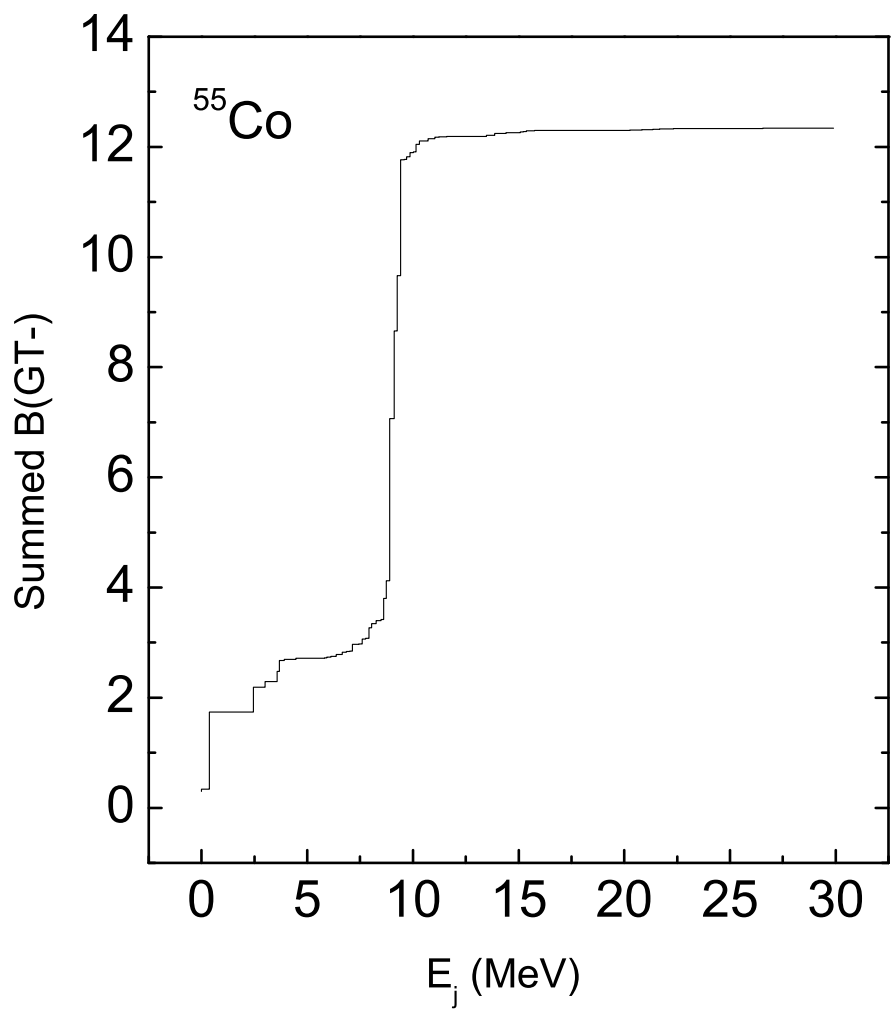}
\end{tabular}
\end{center}
\end{figure}
\end{document}